\documentclass[%
 reprint,
 amsmath,amssymb,
aps,
prl
]{revtex4-1}
 
\usepackage{natbib} 

\usepackage{graphicx}

\usepackage{hyperref}

\hypersetup{
pdfauthor={Piotr Gniewek, Bogumi{\l} Jeziorski},
pdftitle={Exchange interaction energy}
}

\newcommand*{\la}{\langle}
\newcommand*{\ra}{\rangle}

\begin{document}

\title{Determination of the exchange interaction energy \\
   from the polarization expansion of the wave function}

\author{Piotr Gniewek}
\email[]{pgniewek@tiger.chem.uw.edu.pl}
\author{Bogumi\l{} Jeziorski}
\email[]{jeziorsk@chem.uw.edu.pl}
 
\affiliation{Faculty of Chemistry, University of Warsaw, 
Pasteura 1, 02-093 Warsaw, Poland}

\date{\today}

\begin{abstract}
The exchange  contribution to the energy of the hydrogen atom interacting 
with a proton is calculated from the polarization expansion of the 
wave function using the conventional surface-integral formula and two 
formulas involving volume integrals: the formula of the symmetry-adapted 
perturbation theory (SAPT) and the variational formula recommended by us. 
At large internuclear distances $R$, all three formulas yield the correct 
expression $-(2/e)Re^{-R}$, but approximate it with very different convergence 
rates. In the case of the SAPT formula, the convergence is geometric with 
the error falling as $3^{-K}$, where $K$ is the order of the applied 
polarization expansion. The error of the surface-integral formula 
decreases exponentially as $a^K/(K+1)!$, where 
$a=\ln2 -\tfrac{1}{2}$. The variational formula performs best, 
its error decays as $K^{1/2} [a^{ K}/(K+1)!]^2$. 
These convergence rates are much faster than those resulting from 
approximating the wave function through the multipole expansion. 
This shows the efficiency of the partial resummation of the multipole 
series effected by the polarization expansion. Our results demonstrate also 
the benefits of incorporating the variational principle into 
the perturbation theory of molecular interactions. 
\end{abstract}



\maketitle


It is impossible to understand the world without the knowledge 
of intermolecular interactions \cite{Feynman:13}. Not only do they govern the 
properties of gases \cite{Cencek:12}, liquids \cite{Bukowski:07}, 
and solids \cite{Woodley:08}, but also influence chemical reactivity 
\cite{Tomza:15} and determine the structure of complex biological 
systems \cite{Fiethen:08}.

The most straightforward perturbation treatment of molecular interactions, 
known as the {\em polarization approximation} \cite{Hirschfelder:67} or 
{\em  polarization expansion}, consists in an application of the standard 
Rayleigh-Schr\"odinger perturbation theory, with the zeroth-order 
Hamiltonian $H_0$ taken as the sum of the non-interacting monomer 
Hamiltonians, and the perturbation $V$ (the interaction operator) 
defined as $V=H-H_0$, where $H$ is the electronic Hamiltonian of the system.  
Polarization expansion  provides the correct, valid for all intermolecular 
distances $R$, definitions of the electrostatic, induction, and dispersion 
contributions to the interaction energy \cite{Jeziorski:94}. 
It is well known, however, that in a practically computable finite order, 
the polarization expansion for the energy is not able to recover the exchange 
energy, the basic repulsive component of the interaction potential 
that determines the structure of molecular complexes and solids. 
It is also known \cite{Ahlrichs:76, Jeziorski:82} that the polarization 
series provides the asymptotic expansion of the {\em primitive function} 
$\Phi$ \cite{Hirschfelder:67}, 
\begin{equation}\label{Phi}
   \Phi = \varphi^{(0)} + \varphi^{(1)} 
   + \cdots + \varphi^{(K)} + O( R^{-\kappa(K+1)} ) , 
\end{equation}
where $\varphi^{(k)}$ is the $k$th-order (in $V$)  polarization correction 
to the wave function and $\kappa=3$ for interactions of neutral monomers, 
and $\kappa=2$ when at least one of the monomers is charged. 
Equation~(\ref{Phi}) represents the {\em genuine}  primitive function in the 
sense of Kutzelnigg~\cite{Kutzelnigg:80}, i.e., the function which, after 
appropriate symmetry projections ${\cal A}_{\nu}$, yields correctly all 
asymptotically degenerate wave functions $\Psi_{\nu}$ of the interacting 
system, ${\cal A}_{\nu}\Phi =\Psi_{\nu}$, and which is localized in the same 
way as the zeroth-order wave function $\varphi^{(0)}$.  
Using the exact wave  functions $\Psi_{\nu}$, Eq.~(\ref{Phi}) can be written 
in an equivalent, mathematically more precise form \cite{Jeziorski:82} 
\begin{equation}\label{eq:Psi_1}
   |\!| \Psi_{\nu}  - {\cal A}_{\nu} \Phi^{(K)} |\!| = O( R^{-\kappa(K+1)} ) , 
\end{equation}
where $\Phi^{(K)} =\varphi^{(0)} + \varphi^{(1)} +\cdots  + \varphi^{(K)}$ is 
the polarization function through the $K$th order and $|\!|\cdot|\!|$ is the 
usual ${\cal L}^2$ norm.

While methods of calculating the large-$R$ asymptotic behavior of the 
polarization energies (electrostatics, induction, dispersion) are well 
developed and there is a great deal of information about the corresponding 
asymptotic constants \cite{Jeziorski:94}, very little is known about 
the asymptotic behavior of exchange energy. Even the functional form of 
its asymptotic decay for system as simple as two 
hydrogen atoms still stirs controversy \cite{Gorkov:64,Herring:64,Burrows:12}. 
The reason of the difficulty is that the exchange energy, as the result of 
the resonance tunneling of the electrons between the Coulomb wells of 
the interacting atoms,  is sensitive to the wave function values in the 
classically forbidden region of multidimensional configuration space. 
The conventional, basis set based methods of electronic structure theory are 
not well suited to accurately model the wave function in this region.

Only for the interaction of the hydrogen atom with a proton, 
i.e., for the H$_2^+$ system, the asymptotic expansion of the exchange energy 
is known from the tour de force study of Refs.~\cite{Damburg:84, Cizek:86}. 
For this system the exchange energy $J(R)$ is defined as $J=(E_g-E_u)/2$, 
where $E_g$ and $E_u$ are the energies of the lowest \emph{gerade} 
and \emph{ungerade} states of the Hamiltonian  
$H= -\Delta/2 - r_a^{-1} -r_b^{-1} +R^{-1}$, $r_a$ and $r_b$ being the 
distances of the electron to the nuclei $a$ and $b$.
Using semiclassical methods the authors of Refs.~\cite{Damburg:84, Cizek:86} 
found that for H$_2^+$ the exchange energy has the following asymptotic 
expansion: 
\begin{equation}\label{eq:J_asymptotics}
    J(R) \sim (2/e)\,R e^{-R}( j_0  +  j_1\, R^{-1}  
    +  j_2 \,R^{-2}  +  \dots ) ,
\end{equation}
where $j_0 = -1, j_1 = -1/2$, etc. 
Atomic units $\hbar$=$m_e$=$e$=1 are used in Eq.~(\ref{eq:J_asymptotics}) 
and throughout the paper.

In this work we shall consider three formulas expressing $J(R)$ in terms of 
$\Phi$. The physical picture of electrons tunneling from one potential well 
to the other is reflected by the surface-integral 
formula~\cite{Firsov:51,Holstein:52,Herring:62}. Using the notation appropriate 
for H$_2^+$ this formula takes the form 
\begin{equation}\label{eq:J_surf_1}
    J_{ \rm surf } [ \Phi ] = \frac{ \int_M \Phi \nabla \Phi d \mathbf{S} }{
    \la \Phi | \Phi \ra  - 2 \int_{ \rm right } \Phi^2 d V } , 
\end{equation}
where $M$ is the plane perpendicular to the bond axis passing through the 
center of the molecule and the volume integral with subscript ``right'' is 
taken over that half of the space restricted by $M$ where the function $\Phi$ 
is not localized. 
Surface integrals, which are cumbersome in the case of many-electron systems, 
can be avoided if one uses volume-integral formulas: the so-called SAPT formula 
\cite{Gniewek:14}, employed in symmetry-adapted perturbation theory (SAPT) 
\cite{Jeziorski:80,Cwiok:92:SRS}, and the variational formula recommended 
recently by the present authors \cite{Gniewek:15}. In the notation specified 
for H$_2^+$  these formulas have the form: 
 \begin{equation}\label{eq:J_SAPT_1}
    J_{ \textrm{SAPT} } [ \Phi ] = \frac{ 
    \la \varphi^{(0)} | V {\cal P}  \Phi \ra  \la \varphi^{(0)} | \Phi \ra 
    - \la \varphi^{(0)} | V \Phi \ra \la \varphi^{(0)} | {\cal P}  \Phi \ra 
    }{ \la \varphi^{(0)} | \Phi \ra^2 
    - \la \varphi^{(0)} |  {\cal P} \Phi \ra^2 } , 
\end{equation}
\begin{equation}\label{eq:J_var_1}
    J_{ \textrm{var} } [ \Phi ] = \frac{ 
    \la \Phi | H  {\cal P}  \Phi \ra \la \Phi | \Phi \ra 
    - \la \Phi | H \Phi \ra \la \Phi |  {\cal P}  \Phi \ra 
    }{   \la \Phi | \Phi \ra^2 - \la \Phi |  {\cal P} \Phi \ra^2 } , 
\end{equation}
where $ {\cal P}$ denotes the operator inverting the electron coordinates with 
respect to the center of the molecule.

A direct calculation of the primitive function $\Phi$ without a prior knowledge 
of $\Psi_{\nu}$ is very difficult. 
In principle $\Phi$ can be obtained using the Hirschfelder-Silbey (HS) 
perturbation expansion \cite{Hirschfelder:66}, which quickly converges 
for H$_2^+$ \cite{Chalasinski:77} and leads to very accurate values of the 
exchange energy when formulas (\ref{eq:J_surf_1}) and (\ref{eq:J_SAPT_1}) are 
evaluated with the converged $\Phi$ \cite{Gniewek:14}.  
However, the HS theory is not feasible for many-electron systems and we have 
at our disposal only asymptotic approximations to $\Phi$, given by the 
multipole series for the wave function \cite{Ahlrichs:76,Jeziorski:94} or by 
the polarization expansion of Eq.~(\ref{Phi}). The analytic study for H$_2^+$ 
has shown \cite{Gniewek:16} that the multipole expansion of $\Phi$, when 
inserted in Eqs.~(\ref{eq:J_surf_1})-(\ref{eq:J_var_1}), predicts correctly 
the leading $j_0$ term in Eq.~(\ref{eq:J_asymptotics}) but the convergence 
to the exact result is slow (harmonic) when the SAPT formula is used and 
geometric with the ratio of 1/2 and 1/4 when the surface-integral and 
variational formulas are used, respectively.

In the present work we show the results that one obtains using the 
polarization expansion for $\Phi$, i.e., the results of evaluating 
Eqs.~(\ref{eq:J_surf_1})-(\ref{eq:J_var_1}) with the function $\Phi^{(K)}$. 
Since the perturbation $V$ has the  infinite multipole expansion, each 
polarization correction $\varphi^{(n)}$ accounts for the interaction of 
infinitely many multipoles. The polarization expansion includes not only the 
charge-overlap effects \cite{Jeziorski:94} but may also be viewed as 
a selective, infinite-order resummation of the multipole expansion. One can 
expect, then, that the polarization expansion of the wave function can give 
better approximation to the exchange energy than the multipole expansion.

\vspace{1ex}

\emph{Wave function asymptotics.} 
The polarization corrections to the wave function, referred for brevity as 
\emph{polarization functions}, are defined by the recurrence relations 
\begin{equation}\label{eq:phi_RS_corrections}
    ( H_0 - E_0 ) \varphi^{(k)} = - V \varphi^{(k-1)} 
    + \sum_{m=1}^k E^{(m)} \varphi^{(k-m)} 
\end{equation}
where $E^{(k)} = \la\varphi^{(0)}|V\varphi^{(k-1)}\ra$ and the ground-state 
of the hydrogen atom $a$ is taken as the zeroth-order approximation, 
i.e., $\varphi^{(0)} = \pi^{-1/2}e^{-r_a}$, $E_0 = -1/2$.

In our previous work \cite{Gniewek:15}, we showed that the asymptotics 
of $J(R)$, i.e., the value $j_0$ of Eq.~(\ref{eq:J_asymptotics}), when 
calculated from Eqs.~(\ref{eq:J_surf_1})-(\ref{eq:J_var_1}) depends only on 
the values of $\Phi$ on the line joining the nuclei. Thus, if the polarization 
function $\varphi^{(k)}$ is written as $\varphi^{(0)} f^{(k)}(r_a,\theta_a)$, 
where $\theta_a$ is the angle at nucleus  $a$ in the triangle formed by the 
nuclei and the electron, then the angular dependence of $f^{(k)}(r_a,\theta_a)$ 
does not affect the value of $j_0$ and the function $f^{(k)}(r_a,\theta_a)$ can 
be replaced by its value at $\theta_a =0$, i.e., by $f^{(k)}(r_a,0)$. 
We have shown \cite{Gniewek:16} that in the large-$R$ asymptotic expansion of 
$f^{(k)}(r_a,0)$, 
\begin{equation}\label{Fkas}
f^{(k)}(r_a,0) \sim \sum_n R^{-n} \sum_{m=0}^{n} t^{(k)}_{nm}\, r_a^m , 
\end{equation}
only the dominant, $m=n$ terms contribute to the asymptotics of $J(R)$. Thus, 
in calculating this asymptotics, $f^{(k)}(r_a,0)$ can be replaced by 
the simpler function 
\begin{equation}\label{eq:fk_1}
    \widetilde f^{(k)}(r_a) \sim \sum_{n} t_{nn}^{(k)} ( r_a / R )^n .
\end{equation}
In Ref.~ \cite{Gniewek:16} we have shown that the coefficients 
$t^{(k)}_{n}\equiv t^{(k)}_{nn}$ in Eq.~(\ref{eq:fk_1}) satisfy the recurrence 
relation 
\begin{equation}\label{eq:tkn_1}
    t_{n}^{(k)} = \frac{1}{n} \sum_{j=2k-2}^{n-2} t_{j}^{(k-1)} 
\end{equation}
with the initial $k=0$ values given by $t^{(0)}_n =\delta_{n0}$ (we assume that 
a sum is zero when the lower summation limit exceeds the upper one). 
Although the closed-form expression for $t_n^{(k)}$ is unknown, one can show 
that the series of Eq.~(\ref{eq:fk_1}) converges for $r_a < R$ (hence on 
the line joining the nuclei) to the expression 
\begin{equation}\label{eq:fk_tkn}
    \widetilde f^{(k)}(r_a) 
    = \big [ -r_a / R - \ln \big ( 1 - r_a / R \big ) \big ]^k / k! . 
\end{equation}
Eq.~(\ref{eq:fk_tkn}) means that $g^{(k)}(z) =[-z-\ln(1-z)]^k/k!$ \
is the generating function of $t^{(k)}_n$. To prove this it is sufficient 
to note that $g^{(k)}(z)$ satisfies the equation 
\begin{equation}\label{gkz}
   \frac{d}{dz} g^{(k)}(z) = \frac{z}{1-z}\, g^{(k-1)}(z), 
\end{equation}
expand both sides of Eq.~(\ref{gkz}) in powers of $z$, and compare coefficients 
at $z^n$. 
Note that the series of functions $\widetilde f^{(k)}(r_a)$ converges to
$e^{-r_a/R} / ( 1 - r_a/R )$, the function obtained earlier via the WKB method 
\cite{Holstein:52,Herring:62} and shown to represent the dominant 
contribution to the infinite-order polarization function \cite{Scott:91}. 
Thus, our results are consistent with the findings of Ref.~\cite{Scott:91}.

\vspace{1ex}

\begin{figure}[t!]
   \includegraphics{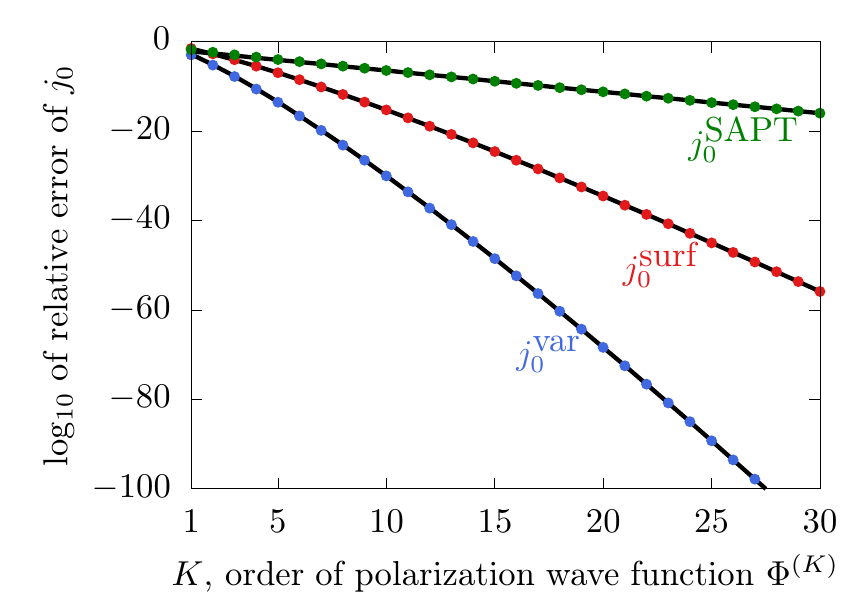}
   \caption{\label{figure} (Color online) Logarithms of errors of $j_0$ 
   calculated using the variational, SAPT, and surface-integral formulas. 
   Dots -- calculated values, solid lines -- the error estimates 
   of Eqs.~(\ref{error_surf}), (\ref{error_var}), and (\ref{3-K}).  }
\end{figure}

\emph{Surface-integral formula.} 
We shall denote by $j_0^{\rm surf}[\Phi^{(K)}]$, $j_0^{\rm SAPT}[\Phi^{(K)}]$, 
and $j_0^{\rm var}[\Phi^{(K)}]$ the approximations to $j_0$ obtained when the 
polarization function $\Phi^{(K)}$ is used in the surface-integral, SAPT, and 
variational formulas, Eqs.~(\ref{eq:J_surf_1})-(\ref{eq:J_var_1}), 
respectively. 
Tang \emph{et al.} \cite{Tang:91} showed that the asymptotics 
of $J_{\rm surf}[\Phi]$ can be determined from the expression 
$-Re^{-R} [F(R/2,0)]^2/2$, where the function $F(r_a,\theta_a)$ 
is defined by the factorization $\Phi = \varphi^{(0)} F(r_a,\theta_a)$. 
Approximating $F(r_a,0)$ by the asymptotics of its $K$th-order polarization 
expansion we find 
\begin{equation}
\label{eq:j0_surf_1}
    j_0^{\rm surf}[\Phi^{(K)}] =
    -\frac{e}{4} \bigg [ \sum_{k=0}^K \widetilde f^{(k)} 
       \! \big (\tfrac{R}{2}) \bigg]^2 
    = -\frac{e}{4} \bigg ( \sum_{k=0}^K \frac{a^k}{k!} \bigg)^2 , 
\end{equation}
where $a=\ln2-\tfrac{1}{2}\approx0.19$. Equation~(\ref{eq:j0_surf_1}) has been obtained 
in Ref.~\cite{Tang:91} using a different derivation.  
The correct value of $j_0$ is recovered by the $K \rightarrow \infty$ limit 
of $j_0^{\rm surf}[\Phi^{(K)}]$ equal to $-e^{2a+1}/4 = -1$. 
Furthermore, the error of $j_0^{\rm surf}[ \Phi^{(K)} ]$ decreases rapidly, as 
\begin{equation}\label{error_surf}
    j_0 - j_0^{\rm surf}[\Phi^{(K)}] = -\sqrt{e} \frac{a^{K+1}}{(K+1)!} 
    + O\bigg(\frac{ a^{K+2}}{(K+2)!}\bigg) , 
\end{equation}
in the same way as the truncation error of the exponential series. 
Figure ~\ref{figure} shows the accuracy of Eq.~(\ref{error_surf}).

\vspace{1ex}

\emph{Variational formula.} 
Since $\la\Phi^{(K)}|\Phi^{(K)}\ra=1+O(R^{-4})$ and 
$\la\Phi^{(K)}|H\Phi^{(K)}\ra=E_0+O(R^{-4})$, the coefficient  
$j_0^{\rm var}[\Phi^{(K)}]$ can be extracted from the expression 
\begin{equation}
    J_{\rm var}^{*}[\Phi^{(K)}] 
    = \la \Phi^{(K)} | ( H - E_0 ){\cal P} \Phi^{(K)} \ra . 
\end{equation}
Writing $\Phi^{(K)} = \varphi^{(0)} F^{(K)}$ one can  
show that $j_0^{\rm var}[\Phi^{(K)}]$ can be obtained from 
even simpler formula: 
\begin{equation}\label{eq:J_var**}
    J_{\rm var}^{**}[\Phi^{(K)}] 
    = \bigg \la {\cal P} \varphi^{(0)} F^{(K)} \bigg | \varphi^{(0)} 
    \bigg ( \frac{ \partial }{ \partial r_a } F^{(K)} + V F^{(K)} 
    \bigg ) \bigg \ra ,
\end{equation}
in which the Laplacian of $F^{(K)}$ was neglected since it does not contribute 
to $j_0^{\rm var}[\Phi^{(K)}]$.

Approximating $F^{(K)}$ by $\widetilde F^{(K)} = \widetilde f^{(0)} + 
\widetilde f^{(1)} + \dots + \widetilde f^{(K)}$ 
and noting that $\partial \widetilde F^{(k)} / \partial r_a 
+ V \widetilde F^{(K)} = \widetilde f^{(K)}$, 
cf.~Eq.~(\ref{gkz}), one can represent the asymptotics 
of $J_{\rm var}^{**}[\Phi^{(K)}]$ in terms of integrals 
\begin{equation}\label{eq:bracket_f1_f2}
    \big \la {\cal P} \varphi^{(0)} \widetilde f^{(k_1)} \big 
    | V \varphi^{(0)} \widetilde f^{(k_2)} \big \ra \! 
    = \! - \frac{ R e^{-R} }{ 4 k_1! k_2! } L(k_1,k_2) 
    \! \bigg [ 1 + O \bigg ( \! \frac{1}{R} \! \bigg ) \! \bigg ] 
\end{equation}
where  
\begin{equation}\label{Lk1k2}
    L(k_1,k_2) = \int_{-1}^1 d \eta \ ( 1 + \eta )^2 
    [ \gamma(\eta) ]^{k_1} [ \gamma(-\eta) ]^{k_2} , 
\end{equation}
and $\gamma(\eta) = (\eta-1)/2 + \ln 2 -\ln(\eta+1)$. 
Eqs.~(\ref{eq:bracket_f1_f2}) and~(\ref{Lk1k2}) follow from the integration in 
the elliptic coordinates, $\xi = (r_a + r_b)/R, \eta = (r_a - r_b)/R$, and 
the integration by parts procedure of Eq.~(29) in Ref.~\cite{Gniewek:15}.

Using Eqs.~(\ref{eq:J_var**}) and (\ref{eq:bracket_f1_f2}) one obtains 
\begin{equation}\label{eq:j0var_0}
    \frac{2}{e} j_0^{\textrm{var}} [ \Phi^{(K)} ] 
    = - \frac{1}{4K!} \sum_{k=0}^K \frac{ L(k,K) }{ k! } . 
\end{equation}
For $K=1$ one finds 
\begin{equation}
 \frac{2}{e} j_0^{\textrm{var}}[\Phi^{(1)}] 
 = -\frac{1}{4} \big [  L(0,1) + L(1,1) \big ] 
 = - \frac{989}{540} + \frac{\pi^2}{9} , 
\end{equation}
in agreement with Ref.~\cite{Chipman:73}. 
For arbitrary $K$ Eq.~(\ref{eq:j0var_0}) can be rewritten as 
\begin{equation}\label{ML}
    \frac{2}{e} j_0^{\textrm{var}} [ \Phi^{(K)} ] 
    = -\frac{1}{4K!} \bigg [ M(K) - \sum_{p=1}^\infty T^K_p \bigg ] , 
\end{equation}
where 
\begin{equation}\label{eq:TKp}
    T_p^K =  L(K+p, K) / (K+p)! , 
\end{equation}
and
\begin{equation}\label{MK}
    M(K) = \sum_{k=0}^\infty \frac{ L(k,K) }{ k! } 
    = \frac{ 8 }{ e } \int_0^\infty x^K e^{-x} d x = \frac{ 8 K! }{ e } . 
\end{equation}
To derive Eq.~(\ref{MK}) we changed the order of summation and integration, 
collapsed the exponential series, and used the variable change 
$x = \gamma(-\eta)$.
Since the second-term in the square brackets on the r.h.s. of 
Eq.~(\ref{ML}) vanishes when $K\rightarrow \infty$, 
cf. Eqs.~(\ref{eq:est_sum})~and~(\ref{eq:Laplace's}), 
we see that $j_0^{\textrm{var}} [ \Phi^{(K)} ]$ converges
to the correct value $j_0 = -1$.

\vspace{1ex}

\emph{Variational formula --- the convergence rate.}
For large $K$ and $p=1,2$ 
the integrals $L(K+p,K)$ of Eq.~(\ref{eq:TKp}) 
can be approximated using the Laplace's method \cite{Bender:99}. 
To this end we rewrite them as 
\begin{equation}\label{intKp}
    L(K+p, K) = \int_{-1}^1 (1+\eta)^2 [ \gamma(\eta) ]^p 
    e^{ K \lambda(\eta) } d \eta 
\end{equation}
where $\lambda(\eta) = \ln \big [ \gamma(\eta) \gamma(-\eta) \big ]$. 
As $\lambda(\eta)$ has a single maximum at $\eta = 0$,
for large $K$ only $\eta \in [ - \epsilon , \epsilon  ]$ with a small 
$\epsilon$ contribute significantly to Eq.~(\ref{intKp}). 
Approximating $\lambda(\eta)$ for $|\eta|<\epsilon$ by  the Taylor expansion,
$\lambda(\eta) = \lambda_0 + \lambda_2 \eta^2 + O \big ( \eta^3 \big )$ 
converts Eq.~(\ref{intKp}) into the Gaussian integral, 
see Ref.~\cite{Bender:99} for details,  
\begin{equation}\label{eq:Laplace's}
    L(K+p, K) = \sqrt{ \pi / ( K | \lambda_2 | ) } 
    \,\, a^{2K+p} \big [ 1 + O(K^{-1}) \big ] , 
\end{equation}
where $\lambda_2 = (4a-1)/(4a^2)$.

We shall now estimate the contribution of the subsequent $p>2$ terms of the 
residual series in Eq.~(\ref{ML}). The integrals $L(K+p,K)$ can be bounded 
by the Schwartz inequality 
\begin{equation}\label{eq:Schwartz}
    L(K+p,K) \leq ( P_{2p} Q_{2K} )^{1/2} , 
\end{equation}
where, using again Laplace's method, 
\begin{equation}
    Q_m \equiv \!\! \int_{-1}^1 \! e^{ m\lambda(\eta) } d\eta 
    = \sqrt{ \pi/ ( m |\lambda_2| ) } \, a^{2m} 
    \bigg [ 1 + O \bigg ( \! \frac{1}{m} \! \bigg ) \! \bigg ] 
\end{equation}
and, using the variable change $t = \ln 2 -\ln(t+1)$, 
\begin{equation}
    P_m \equiv \int_{-1}^1 (1+\eta)^4 [ \gamma(\eta) ]^m d\eta 
    = 32 \int_0^\infty e^{ -5t } \big [ e^{ -t } +t -1 \big ]^m dt . 
\end{equation}
Since $e^{-t} +t -1 \leq t$ for $t \geq 0$, it follows that 
$P(m) \leq (32/5) m!/5^m$, so that $T^K_p \leq \widetilde T^K_p$, where 
\begin{equation}
    \widetilde T^K_p 
    = 4 \big [ \tfrac{2}{5} (2p)! Q_{2K} \big ]^{1/2} \, 5^{-p} / (K+p)! . 
\end{equation}
Since $\widetilde T^K_{p+1} / \widetilde T^K_p \leq 2/5$, 
we can estimate the contributions of the terms with $p \geq 3$ by  
\begin{equation}
    \frac{ 1 }{ T^K_1 }\sum_{p=3}^\infty T^K_p 
    \leq \frac{5}{3} \frac{ \widetilde T^K_3 }{ T^K_1 } 
    = \frac{ D \, K^{1/4} }{ (K+2)(K+3) } \bigg [ 1 + O(K^{-1}) \bigg ] ,  
\end{equation}
where $D = 16 [ ( 1-4a ) / ( 2 \pi ) ]^{ 1/4 } / ( 25 a^{3/2} )$. 
In view of Eq.~(\ref{eq:Laplace's}), $T_2^K \sim a^{2K+2} / (K+2)!$, so that 
finally 
\begin{equation}\label{eq:est_sum}
    \sum_{p=1}^\infty T^K_p = T^K_1 \big [ 1 + O \big ( K^{-1} \big ) \big ] .
\end{equation}
Thus, the error of $j_0^{\rm var}[\Phi^{(K)}]$ is dominated 
by the $p=1$ term in the sum in Eq.~(\ref{ML}), 
\begin{equation}\label{error_var}
    j_0 - j_0^{\rm var} [ \Phi^{(K)} ] 
    = - \frac{ A \ a^{2K+2} }{ \sqrt{K} K! (K+1)! } 
    \bigg [ 1 + O\big ( K^{-1} \big ) \bigg ] , 
\end{equation}
where $A = e \sqrt{\pi} / ( 4 \sqrt{1-4a} )$. 
The rapid fall-off of the error of $j_0^{\rm var}[\Phi^{(K)}]$ can be seen 
in Fig.~\ref{figure}.

\vspace{1ex}

\emph{SAPT formula.}
To obtain $j_0^{\textrm{SAPT}}[ \Phi^{(K)} ] $ it is sufficient 
to consider the following approximation to $J_{\textrm{SAPT}}[\Phi^{(K)}] $,
\begin{equation}
    J_{\textrm{SAPT}}^*[\Phi^{(K)}] 
    = \la \varphi^{(0)} | V {\cal P} \varphi^{(0)} F^{(K)}\ra .
\end{equation}
Approximating $F^{(K)}$ by  the sum of  functions $\widetilde f^{(k)}(r_a)$  
one can represent $j_0^{\rm var}[\Phi^{(K)}]$ in terms of integrals 
$\la \varphi^{(0)} | V {\cal P} \varphi^{(0)}\widetilde f^{(k)}(r_a) \ra$. 
Using Eq. (\ref{eq:bracket_f1_f2}) one obtains 
\begin{equation}\label{jSAPT}
    \frac{2}{e} j_0^{\textrm{SAPT}}[ \Phi^{(K)} ] 
    = -\frac{1}{4} \sum_{k=0}^K \frac{ L(k) }{ k! } , 
\end{equation}
with $L(k) = L(k,0)$. When $K\rightarrow \infty$, the sum on the r.h.s. is 
equal to $M(0)$, so in view of Eq.~(\ref{MK}), 
$j_0^{\textrm{SAPT}}[ \Phi^{(K)}]$ converges to the correct value $j_0=-1$.

To calculate the error of $j_0^{\textrm{SAPT }} [ \Phi^{(K)} ]$ we need 
the integrals $L(k)$, for which the variable change $t = \ln 2 -\ln(t+1)$ gives 
\begin{equation}\label{eq:Lk}
   L(k) = 8\!\! \int_{0}^\infty \!\!\! e^{-3t} \big [ e^{-t} + t -1 \big ]^k dt 
   = \sum_{l=0}^k \frac{ 8 k! e_{k-l}(-l-3) }{ l! (l+3)^{k-l+1} } , 
\end{equation}
where $e_n(x)$ is the exponential sum function, 
i.e. the series of $e^x$ truncated after the $x^{n}/n!$ term. 
The large-$k$ asymptotics of $L(k)$ is given by the first, $l=0$ term in 
the sum in Eq.~(\ref{eq:Lk}). It follows that   
\begin{equation}\label{3-K} 
    j_0 - j_0^{\textrm{SAPT}}[ \Phi^{(K)} ] 
    =  -\frac{ 1 }{ 6 e^2 } \,  3^{-K} +O(4^{-K}) . 
\end{equation}
The error of $j_0^{\textrm{SAPT}}[ \Phi^{(K)} ]$ can be compared to the errors 
of the other two formulas in Fig.~\ref{figure}.

\vspace{1ex}

\emph{Summary and conclusions.} 
By solving analytically the model system of the hydrogen atom interacting 
with a proton we found that all three exchange energy formulas considered 
by us correctly predict the large-$R$ behavior of the exchange energy 
if the primitive function is approximated by the standard polarization 
expansion. The correct limit is however approached with very different 
convergence rates. In the case of the SAPT formula, the convergence 
is geometric with the error decaying as $1/3^K$, where K is the order of 
the applied polarization theory. The convergence of the surface-integral 
formula is exponential, with the error decreasing as $a^K/(K+1)!$, 
where  $a=\ln 2 -1/2$. The best convergence occurs for the variational 
formula, for which the error falls off as $K^{1/2} [a^K/(K+1)!]^2$. 
The observed convergence rates are significantly faster than those 
resulting from approximating the primitive function through the multipole 
expansion \cite{Gniewek:15,Gniewek:16}. To make a meaningful comparison, 
cf. Table~\ref{table}, we note that $\Phi^{(K)}$  and the sum of 
the mulitpole expansion through the 2$K$th order in $1/R$, denoted 
by $\Phi_{2K}$, are both accurate through the (2$K$)th order in $1/R$. 
However, $\Phi^{(K)}$, unlike $\Phi_{2K}$, includes a selective 
infinite order summation of higher $R^{-k}$, $k > 2K$ terms. 
The inspection of Table~\ref{table} shows that this infinite order, 
selective summation is very effective in computing the exchange energy, 
independently of the exchange energy expression employed.

The main conclusion of our investigation is that the exchange energy, 
an electron tunneling effect, can be determined from the knowledge 
of the wave function which reflects only the polarization mechanism 
of interatomic interaction. We have shown that this determination 
is particularly effective when the variational principle is employed 
in the perturbation treatment of molecular interactions. 
We expect that this conclusion is general and applies also 
to interactions of larger systems.

\begin{table}[tb]
\caption{\label{table} Decay rate of the error of the leading term 
of exchange energy calculated using truncated multipole $\Phi_{2K}$ 
(Ref.~\cite{Gniewek:16}) and polarization $\Phi^{(K)}$ series. 
$a = \ln2-\frac{1}{2} \approx 1/5$. }
\centering
\begin{tabular}{ l c c  } 
\hline\hline
  \rule{0pt}{0.4cm} & $\Phi=\Phi_{2K}$ & $\Phi=\Phi^{(K)}$ \\[0.5ex]
\hline
\rule{0pt}{0.6cm} $j_0^{\textrm{surf}}[\Phi]$ & $\displaystyle \frac{1}{4^K}$  
      & $\displaystyle \frac{a^K}{(K+1)!}$ \\
\rule{0pt}{0.7cm} $j_0^{\textrm{SAPT}}[\Phi]$ & $\displaystyle \frac{1}{K^2}$  
      & $\displaystyle \frac{1}{3^K}$ \\
\rule{0pt}{0.8cm} $j_0^{\textrm{var}}[\Phi]$  & $\displaystyle \frac{1}{16^K}$ 
      & $\displaystyle \frac{a^{2K}}{K!(K+1)! \sqrt{K}}$ \\[2.5ex]
\hline\hline
\end{tabular}
\end{table}

\begin{acknowledgments}
This work was supported by the National Science Centre, Poland, 
project number 2014/13/N/ST4/03833. 
\end{acknowledgments}

\bibliography{ref}

\end{document}